# Data Driven Charge Transfer Atlas Provides Topological View of Electronic Structure Properties for Arbitrary Proteins Complexes

Fang Liu[1], Hongwei Wang[1], Dongju Zhang[2], Likai Du[1]*,

[1]Hubei Key Laboratory of Agricultural Bioinformatics, College of Informatics, Huazhong Agricultural University, Wuhan, 430070, P. R. China

[2]Institute of Theoretical Chemistry, Shandong University, Jinan, 250100, P. R. China

*To whom correspondence should be addressed.

Likai Du: dulikai@mail.hzau.edu.cn;

**Abstract**

Due to the highly complex chemical structure of biomolecules, the extensive understanding of the electronic information for proteomics can be challenging. Here, we constructed a charge transfer database at residue level derived from millions of electronic structure calculations among 20×20 possible amino acid side-chains combinations, which were extracted from available high-quality structures of thousands of protein complexes. Then, the data driven network (D$^2$Net) analysis was proposed to quickly identify the critical residue or residue groups for any possible protein complex. As an initial evaluation, we applied this model to scrutinize the charge transfer networks for two randomly selected proteins, which highlighted the most critical residues with large node degrees as network hubs. This work may provide us a promising computational protocol for topologically understanding the electronic structure information in the growing number of high-quality experimental proteins structures, with minor computational costs.

**Introduction**

Charge transfer reactions take place in a wide range of biological processes, including photosynthesis, respiration, and signal transduction of biology, enzymatic reactions, gene replication and mutation and so on.[1-5] In biological systems, electron or hole transfer reaction can occur between donors and acceptors separated by a long distance, for example across protein-protein complexes.[6-11] Superexchange theory (or electron tunneling) and the hopping model are commonly used to describe electron and hole transfer processes.[7, 12-15] However, the issue of charge transfer in the entire proteomics is still intriguing and challenging due to the complicated structures of realistic proteins.[16-19]

In recent years, the growing amount of high-quality experimental (X-ray, NMR, cryo-EM) structures have opened space to improve our theoretical understanding of biological charge transfer reactions in the foreseen big data scenario. The building blocks of proteins are only the twenty L-amino acids, which are distinguished by their different side-chain structures and chemical compositions. However, the physical interactions among residues at the local and overall level are quite complicated in contrast to periodic material systems. Bioinformatics scientists have paid much effort to the classification of protein structures in the last decades. And a large number of biological databases were constructed to depict the structural significance of protein complexes.[20-27] In parallel to these exciting developments in structural biology and bioinformatics, it becomes increasingly important to incorporate our available structural knowledge, such as the significance of amino acid preference in proteins, into our physical understanding of charge transfer reactions.

It is interesting to capture the electron or hole transfer chain or network in a human-accessible, topological picture.[28] The complex network analysis is an appeal and popularity to obtain qualitative insights, which has been widely used in various fields of chemical and biological studies[29-32], i.e. protein/protein interactions[33-35], identification of targets for drugs[36-38], chemical reaction network[39-40], metabolic engineering[41-43] and so on. There are also many works discussing how to represent the electron or hole transfer pathways connecting electron donating and accepting cofactors in biological[16, 44-49] and disordered material systems[50-54]. For example, Beratan et. al. suggested to use the graph theory to search and identify tunneling pathways or pathway families in biomolecules.[44-45, 49] However, the electronic couplings were empirical in their work, for which electronic couplings along a given pathway were written as a product of a hypothetical closest contact terms, involving covalent, hydrogen bond, and van der Waals interactions. Nevertheless, such simple coarse-grained models remain of considerable interest for exploring charge transfer in biological systems[49].

The charge transfer electronic coupling parameter is an important component for biological charge transfer reaction, which can be derived from various empirical or semi-empirical models[16, 44-45, 49, 55-57] and from direct electronic structure calculations.[58-63] Nowadays, in the era of high performance computing, the sophisticated models are becoming increasingly possible to obtain these electronic couplings for ensembles of structures in biomolecules. Therefore, it is possible to directly derive the charge transfer coupling parameters for millions of molecular fragments, which sufficiently represent most possible occurrences in proteins database.

In this work, we propose the data driven network ($D^2$Net) analysis tools to obtain the topological charge transfer features in any possible protein complex. First, we presented a computational protocol to construct the charge transfer database, which provided an overall view of charge transfer coupling atlas among millions of amino acid side-chain conformations. The sophisticated charge transfer parameters in the database could be reused with minor computational costs. Thus, this charge transfer database as a powerful look-up table was applied to construct complex charge transfer networks for realistic protein complexes. This is also our first attempt to simulate charge transfer networks in realistic proteins by incorporating sufficient structural information. The global topology highlights the most critical residues in charge transfer reactions act as network hubs. In spite of its simplifications, the complex network analysis shows the unique ability to place different charge transfer mechanisms on the same footing under a reasonable graph metric.

## 2. Methods and Theoretical Details

### 2.1 Mega Data Sets for the Charge Transfer Atlas

The initial step to develop any data driven or informatics based model is the data collection procedure. Here, we extracted the structural data set from an improved web version[64] of the "Atlas of Protein Side-Chain Interactions", which were derived from thousands of unique structures of protein complexes solved by X-ray crystallography to a resolution of 2.0 Å or better. As of June 2017, the Atlas comprised 482555 possible amino acid side-chain conformations for 20×20 sets of amino acid contacts. And the snapshot of the database on June 2017 can also be obtained from us upon request.

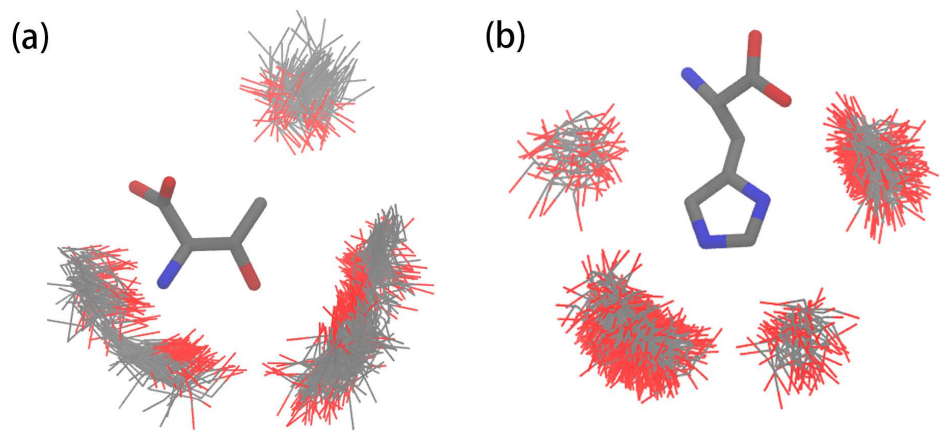

**Figure 1.** Examples for spatial distribution of amino acid side-chain interactions and their associated clusters are visualized. The Thr/Ser (a) and His/Glu (b) pairs are given to illustrate the distinct geometric distributions for each type of amino acid combinations. Note that, the hydrogen atoms are not shown for clarity.

For each amino acid pair, the atlas shows how one amino acid side chain is distributed with respect to the other in the three-dimensional space. The preferred interaction geometric patterns are revealed by clusters in the distributions. As shown in Figure 1, the amino acid pairs have preferred interaction patterns, indicating their packing is not entirely random.[66-68] The three atoms for the central amino acid are used to define the frame of reference (Figure S1), which is summarized in the supporting information. Each dimer, consisting of one amino acid side-chain pair is transformed to utilize the same frame of reference, with respect to one amino acid, which yields 20 distributions of amino acid residues around each one. The procedure to extract each dimer has been described in the work of Singh and Thornton,[65] for which the clustering algorithm can be summarized as follows.

The root-mean-square deviation (RMSD) between atom positions was calculated for all pairs of amino acid side-chains using the same frame of reference. Any side chain with an RMSD of less than 1.5 Å from the selected side-chain is considered a "neighbor". The side-chain structure with the largest number of neighbors is taken to be the cluster representative. This side-chain and all its neighbors form the largest cluster, which are then removed from the original data set. And the calculation is repeated to obtain the cluster representative for the second largest cluster, etc. The clustering procedure was repeated up to six times for each type of amino acid dimer, depending on the size of the geometric cluster. For each cluster, the most representative side-chain is the one with the minimum RMSD to all of the other side-chains in the same cluster.

**2.2 Charge Transfer Couplings from the Bio-molecules Tight Binding Method**

The tight binding model is an effective approach to study complicated molecular systems. The single-electron motion equation for biomolecule systems can be derived according to the idea of tight-binding approximation. Here, we only provide a brief introduction of the tight binding method for biomolecules (bioTB), following the previous work of Liu and co-workers[69-71]. Further theoretical details are given in the supporting information.

The building blocks of bio-molecules are just some repeated structural units, i.e. amino acids for proteins, nucleotides for DNA and *so on*, and the many-electron Hamiltonian in this physical view can be expressed as,

$$H = \sum_i -\frac{1}{2}\nabla_i^2 + \sum_{i,L}\sum_{a \in L} -\frac{Z_a}{r_{ai}} + \frac{1}{2}\sum_{i,L}\sum_{j \in L, j \neq i}\frac{1}{r_{ij}} + \sum_L \sum_{a,b \in L, a<b}\frac{Z_a Z_b}{R_{ab}} + \sum_{L,M}\sum_{a \in L, b \in M}\frac{Z_a Z_b}{R_{ab}} \quad (1)$$

In Eq. 1, L and M refer to the repeated structural units or sites. The last term refers to the repulsion between the nuclei, which has no effect on the electronic structure of the bio-molecules. The Eq. 1 can be rewritten as the sum of one-electron operators,

$$H = \sum_i \left[ -\frac{1}{2}\nabla_i^2 + \sum_L V_L(i) \right] \quad (2)$$

And

$$V_L(i) = \sum_{a \in L} -\frac{Z_a}{r_{ai}} + \frac{1}{2}\sum_{j \in L, j \neq i}\frac{1}{r_{ij}} \quad (3)$$

The corresponding many-electron Schrödinger equation can be solved *via* the one-electron Schrödinger equation,

$$h_i \psi = \varepsilon_i \psi \quad (4)$$

Whereas, $\psi$ is the one-electron wave function for the bio-molecules, and the one-electron

Hamiltonian can be written as,

$$h_i = -\frac{1}{2}\nabla_i^2 + \sum_L V_L(i) \tag{5}$$

The first step to solve the Eq. 4 is to calculate the one-electron Schrödinger equation of the isolated structural unit at the non-perturbation state.

$$\left[-\frac{1}{2}\nabla_i^2 + V_L\right]\phi_l = \varepsilon_l^0 \phi_l \tag{6}$$

In Eq. 6, $\phi_l$ is the molecular orbital of one structural unit L, and $\varepsilon_l^0$ is the corresponding orbital energy. The Eq. 6 can be solved by available electronic structure methods, such as the Hartree-Fock or DFT method. And the electronic orbital for the entire bio-molecules can be expanded as the linear combination of site orbitals for each repeated structural unit.

$$\psi = \sum_l c_l \phi_l \tag{7}$$

Substituted the above formula into Eq. 4, we can derive the formulas for the on-site energy and transfer integral,

$$\varepsilon_n = \langle \phi_n | h | \phi_n \rangle = \langle \phi_n | -\frac{1}{2}\nabla^2 + \sum_L V_L | \phi_n \rangle \tag{8}$$

$$t_{n,n+1} = -\langle \phi_n | h | \phi_{n+1} \rangle = -\langle \phi_n | -\frac{1}{2}\nabla^2 + \sum_L V_L | \phi_{n+1} \rangle \tag{9}$$

The summation runs over all possible structural units L, however, only the neighboring units are required to be considered in the tight binding approximation. The transfer integral describes the ability to perform the charge transfer among neighbor sites, meanwhile, the on-site energy describes the ability to move or inject an electron from a specific site.

Thus, the on-site energy for unit $n$ only needs the potential information of unit $n$ and its closest neighboring units **C**, and the formula of on-site energy can be simplified as,

$$\begin{aligned}\varepsilon_n &\approx \langle \phi_n | -\frac{1}{2}\nabla^2 + V_n + \sum_{L \in \mathbf{C}} V_L | \phi_n \rangle \\ &= \varepsilon_n^0 + \langle \phi_n | \sum_{L \in \mathbf{C}} V_L | \phi_n \rangle\end{aligned} \tag{10}$$

The transfer integral only requires the potential of structural units $n$ and $n+1$, that is

$$t_{n,n+1} \approx -\langle \phi_n | -\frac{1}{2}\nabla^2 + V_n + V_{n+1} | \phi_{n+1} \rangle \tag{11}$$

Because the tight binding model is corresponding to the orthogonal basis, the Löwdin method[71-73] is performed to minimize the orbital overlap. And the effective transfer integral can be transformed to,

$$t^{eff}_{n,n+1} = \frac{t_{n,n+1} - \frac{1}{2}(\varepsilon_i + \varepsilon_j)s_{n,n+1}}{1 - s^2_{n,n+1}} \quad (12)$$

In Eq. 12, $s$ is the orbital overlap integral between units. This transformation shows minor effects on the on-site energy and therefore can be ignored in our code implementation. And one of the implementation of these formulas can be found in http://github.com/dulikai/bioX.

The on-site energy and transfer integral, as two basic variables, are the diagonal and off-diagonal elements of the tight binding Hamiltonian, which are basic physical variables in the study of DNA damage and respiration, photosynthesis and the design of molecular electronic device and charge transfer problems.

Once the tight binding Hamiltonian is constructed, we may directly solve the well-known eigenvalue equation (**HC=EC**) to obtain the electronic structure information of bio-molecules. However, we suggest that the topological understanding of the electronic properties for proteins can be possible by network analysis, without directly solving such electronic structure equation of bio-molecules, see below.

**2.3 Data Driven Networks for Charge Transfer in Proteins**

The charge transfer network is the generalization of charge transfer chains/pathways, which consists sets of substrates in their reduced and oxidized forms. In order to construct the protein charge transfer network, each residue is represented by a vertex in the graph, and the edge represents the strength of the charge transfer coupling among residues. The charge transfer rate is proportional to the square of electron transfer coupling strengths. Note that, the charge transfer rates depend on electronic coupling elements, reorganization energies, and driving forces. However, the exact evaluation of these contributions in realistic proteins is computationally prohibitive, which also significantly complicates the network analysis. Therefore, it is more straightforward to use the strength of charge transfer couplings in our network analysis.[50, 52-54, 74-75]

As we are primarily concerned with the network topology, the undirected graph is considered with the following adjacency matrix, and the edge can be only possible to be 0 or 1.

$$A_{ij}(\mathbf{x}) = \begin{cases} 1 & \mathbf{x} \in V \\ 0 & \text{else} \end{cases} \quad (13)$$

The value of the edge is assign to be 1, only if the pairwise side-chain structure **x** is classified into a known cluster (**V**) with significant charge transfer strength in the charge transfer atlas. The selection of the threshold value is nontrivial, and we mainly consider this issue on two aspects. First, the threshold value should be relatively larger than the background value or the average value of the entire database. This is because our D²Net model is statistic or informatics based in nature, and we need to filter the noisy data, in order to highlight the most significant features. And the background value of the database is in the range of 0.01~0.02 eV. Second, the threshold value should not be very large to ignore much valuable electronic structure information. After a few attempts in the network analysis, the threshold of significant charge transfer coupling is set to be 0.02 eV in this work.

As shown in Scheme 1, we suggest a procedure to construct the data driven network (D²Net) model by reusing the charge transfer coupling atlas. This procedure is based on our experience on a few model systems[76-77], and we could reasonably estimate the charge transfer couplings for any possible amino acid conformations in realistic proteins, if our protein charge transfer atlas is large enough.

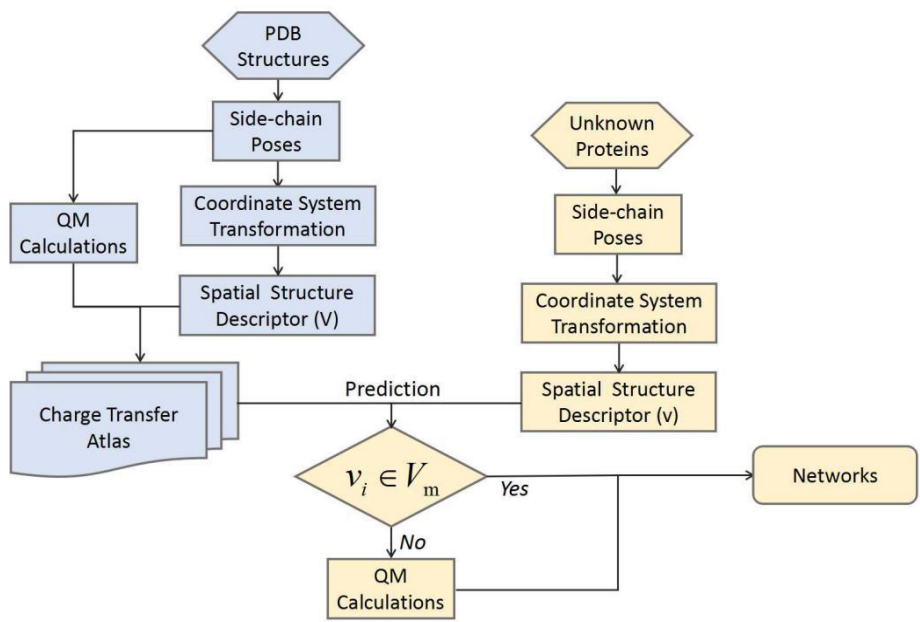

**Scheme 1**. The charge transfer atlas is constructed from thousands of PDB structures, along with QM calculations. Once the charge transfer atlas is constructed, the network model and other statistics can be derived by a variety of means.

In order to predict the geometric dependent fluctuations, one possible problem with our data driven approach is the determination of how far one unknown molecular structure can move away from the data set in geometric space before the estimation approach fails. Actually, we use the relative error to monitor the general distances between an arbitrary structure and a set of known structures. The relative error (RE) is defined as follows,

$$RE(k) = \left| \frac{\mathbf{X} - \mathbf{X}(k)}{\mathbf{X}(k)} \right| \times 100\% \tag{14}$$

In the above equation, $k$ runs over the known molecular structures. And $\mathbf{X}$ and $\mathbf{X}(k)$ are molecular coordinates as vectors, which are used as fingerprints to search the available charge transfer atlas with millions of side-chain pairs. And the unknown dimer structure can be classified to be similar with one of the known structures. If the calculated RE is larger than 20% for any known structures in the known data sets, the assignment for such unknown structure is said to be failed.

Then, an ab initio calculation of the charge transfer coupling is called to deal with the incompleteness of this charge transfer atlas, meanwhile, this unknown structure is added into our charge transfer atlas for future similarity assignment. Therefore, this process is boot-strapped from the available charge transfer atlas along with minor number of ab initio calculations. The entire computational protocol has been automated in a series of Python codes.

As an initial evaluation of the D$^2$Net model, we randomly selected two crystal structures of proteins from the PDB. One is the HIV-1 integrase core domain (PDB code 1qs4), and the other is the complex between the human H-Ras protein and the Ras-binding domain of C-Raf1 (Ras-Raf), which is central to the signal transduction cascade. The starting structures for the simulations of the human unbound proteins and complexes were taken from the PDB database (PDB codes: 1rrb, 121p, 1gua) and modified to achieve consistency with respect to the biological source and the number of amino acids.[78] For demonstrative purposes, the molecular structure was only minimized in the molecular simulations. The mega data sets in our charge transfer atlas are large enough to represent most possible amino acid side-chains orientations in realistic protein structures. And the failure of the predictions is below 0.1% in both systems.

## Results and Discussions

At first, we established a vocabulary to describe how the conformation ensemble influence electronic couplings for millions of amino acid side-chain combinations. Although, previous studies[66-68] have revealed the relative abundance of various modes of amino acid contacts (van der Waals contacts, hydrogen bonds), relatively little is known about the qualitative charge transfer coupling terms of these noncovalent interactions. This database is helpful to unravel the richness of biological charge transfer coupling in realistic proteins, which would evolve within fluctuating biomolecules structures.

In fact, we have calculated the electronic couplings of the frontier orbitals of amino acid pairs, i.e. HOMOs, LUMOs or HOMO/LUMO, under the tight binding approximation. However, a thorough comparison of the electronic couplings among varius frontier orbitals is beyond the scope of this work, but, instead, we roughly classify them into two categories, namely the electron transfer and hole transfer couplings. The electronic couplings between HOMOs represent the hole transfer process, i.e. removing an electron from the HOMO of one amino acid fragment. The electronic couplings between HOMO and LUMO should also be important, which can be analyzed by the similar protocol. Thus, without losing any generality, we will restrict our study to electronic hole transfer couplings between HOMOs of each amino acid side-chain combinations.

The twenty amino acids can be divided into several groups according to the chemical compositions of their side-chains, that are the hydrophobic group (i.e., GLY, ALA, VAL, ILE, MET, and PHE), polar and neutral group (i.e., SER, THR, CYS, TYR, ASN, and GLN), acidic group (i.e., ASP and GLU) and basic group (i.e., LYS, ARG, and HIS). To simplify the statistical representation of the charge transfer atlas, we defined a resolution for the atlas, for which the unsigned charge transfer coupling below a threshold value was assigned to be zero. Therefore, we can provide a global view of charge transfer couplings among twenty natural amino acids (totally 20×20 combinations) under various resolutions from 0.01 eV to 0.1 eV in Figure 2.

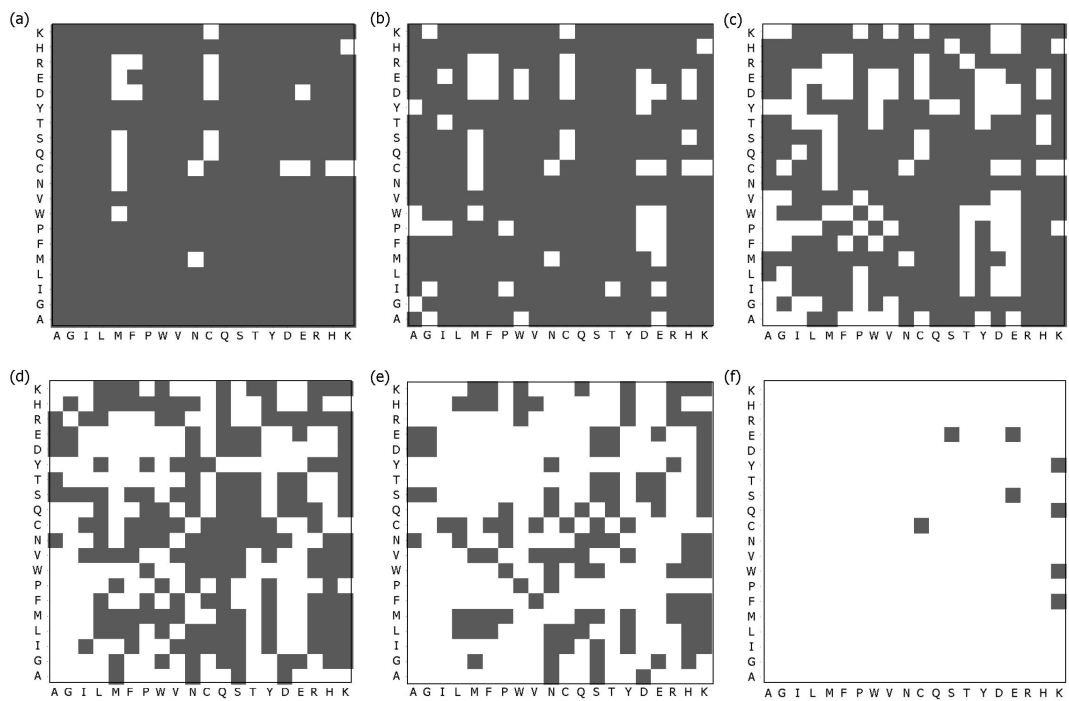

**Figure 2.** The distribution (zero *vs.* one) of unsigned charge transfer integral for each type of amino acid side-chain combinations, averaged over the available structures. The resolution of the heat map is assigned to be 0.01, 0.02, 0.03, 0.04, 0.05 and 0.1 eV for a-f, and each amino acid is referred as one letter. There are only two possible values in the heat map, namely zero and one. The white color and grey color refer to the amino acids below and above the specific resolution.

By analyzing the charge transfer atlas at different resolutions, the first of the important results is that most charge transfer coupling lies below 0.05 eV, however, the charge transfer coupling is not zero in most cases. In general, averaged transfer couplings for the polar/polar or basic/acidic combinations (i.e. K/F, S/E) are greater than the pairwise hydrophobic combinations (i.e. A/G, V/F), with only minor exceptions. Note that, the heat maps are not symmetric because the inhomogeneous of protein structures and the distribution of one type of amino acid in the frame of another reference residue type is distinct. Therefore, we believe this asymmetric feature reflects the significant protein structures which are most probably based on certain geometry preferences between interacting amino acids. This heat map may be helpful as reference materials at hand to qualitatively understand the possible charge transfer feature in proteins.

Then, we tried to describe the charge transfer couplings population for a few selected amino acids pairs in the context of overall geometric distribution. The signed charge transfer couplings distribution for distinct geometric clusters is given in Figure 3. The averaged charge transfer coupling parameters (Figure 2) for each type of amino acid side-chain combinations seem to be not very suitable to describe the distribution of overall geometric clusters. These findings indicate that one must pay close attention when dealing with the geometric ensemble of amino acid pairs in realistic proteins, and the appropriate transfer coupling parameters should be applied only after performing tests on similar geometric features.

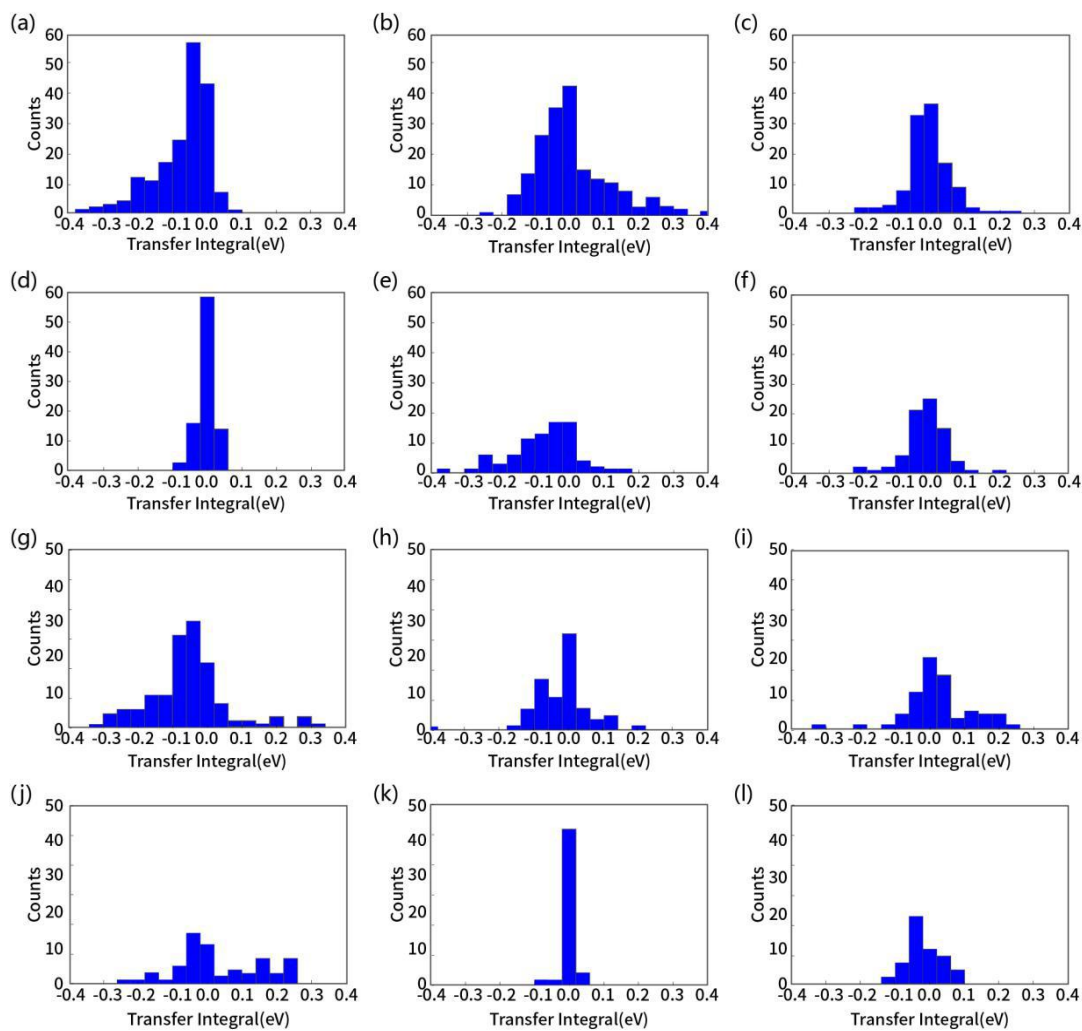

**Figure 3.** The distribution of charge transfer couplings for a few selected amino acids pairs. Six clusters of Lys/Asp combination (a-f) and Ser/Thr (g-l) are shown, whereas Lys and Ser is the center fragment.

Figure 3 also suggests that the charge transfer coupling parameters are found to be very sensitive to the structural orientation of the amino acid pairs. Take the Lys/Asp pair as an example, the charge transfer coupling is very significant between the basic Lys and acidic Asp fragments (Figure 2), however, the distribution varies much from -0.3 eV to 0.3 eV (a-f in Figure 3). The feature of broad distribution is also found for Ser/Thr as one of the polar and neutral combinations. This rather inhomogeneous of electron transfer couplings may be caused by various inter-molecular interactions (covalent, hydrogen bonds, van der Waals, ionic) within realistic proteins.[79-82]. One can therefore conclude that there are no uniform criteria or empirical formula to accurately estimate the electron transfer coupling parameters for a specific amino acid pair, since the board distribution makes the estimation to be validated only for a small fraction of the geometric ensemble.

Next, the protein side-chain charge transfer database is used to construct the electron transfer network for any protein structures, which refers as D$^2$Net model. Because the anisotropic features cannot reasonably be ignored in making estimation of charge transfer couplings among amino acids, we tried to extensively use the charge transfer atlas for our subsequent network analysis. Note that, the network is different from the assumption of a charge transfer chain, for which a given reduced substrate can only transfer its electron to the next substrate in a linear fashion, and in a network model, each reduced substrate is able, in principle, to transfer electrons to any oxidized substrate. The large number of possible residues in proteins leads to charge transfer networks with hundreds or thousands of species, thousands of pathways and an exponential number of possible pathways and mechanisms to be considered. Of course, we may choose to rule out some possible transfer routines, and thus a given network should have a particular topology, for which a charge transfer chain is one specific example.

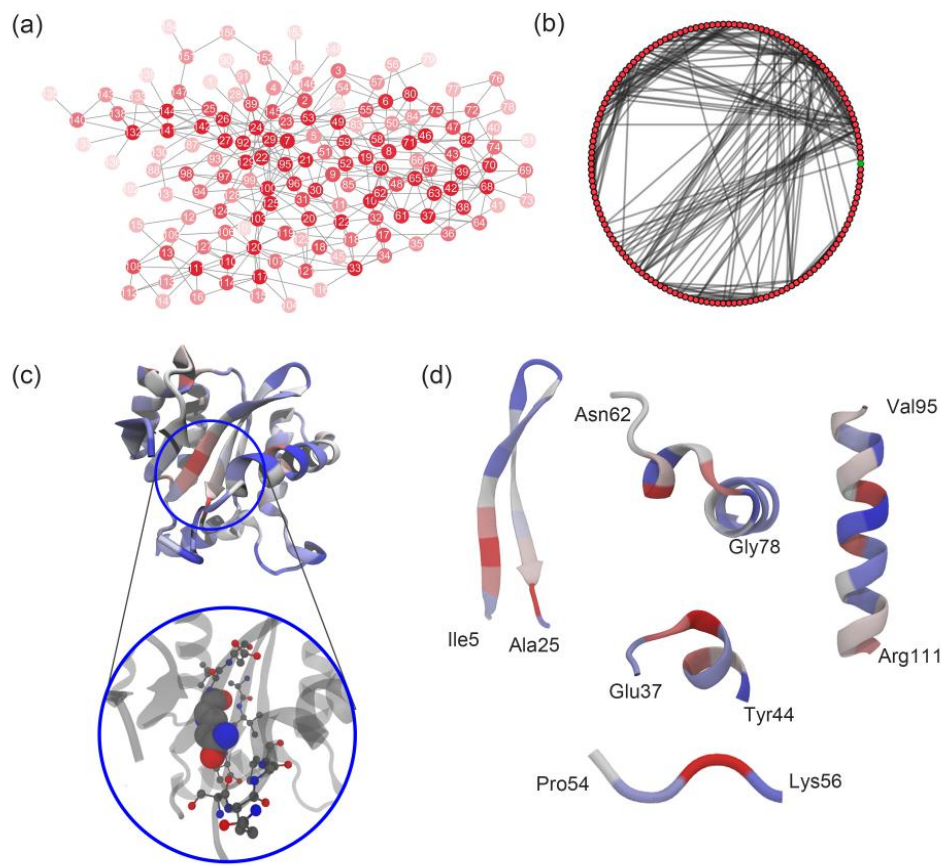

**Figure 4.** Construction of charge transfer network among residues at the resolution of 0.02 eV for the HIV-1 integrase core domain. The graph representations of protein charge transfer network are given with CoSE (a) and circular (b) layout. (c) The node degree distribution is mapped to the protein structure. (d) The secondary structures as building blocks with higher node degree are shown. The red color refers to the "hot" residues with large node degrees, while the blue color means "cold" residue with much smaller node degrees.

Figure 4 shows the charge transfer network analysis of the HIV-1 integrase core domain. To facilitate our following discussions, the words "hot" and "cold" are used to indicate the nodes (residues) with high and low degree in the network. In Figure 4a, we applied the CoSE layout[83] to visualize the topological feature of undirected compound graphs, and this algorithm highlights the most important nodes ("hot" residues) and their surrounding nodes in an intuitive way. The "hot" residues with deep red color may form a reduced version of the protein core i.e. residue number 7, 22, 24, 29 and so on. And the remainder of the "cold" protein residues (light red nodes) does not substantially change transfer coupling when included. In Figure 4b, the circular layout is used to represent alternative view of the protein charge transfer network. The advantage of a circular layout in the biological applications is its neutrality, because none of the nodes (residues) is given a privileged position by placing all vertices at equal distances from each other as a regular polygon. The circular layout views in protein systems may provide useful insights on the topology relationship among the protein structures. In summary, the network representation provides us a way to understand the topology of charge transfer networks.[28] These derived results may be helpful for the mutation experience to improve the electronic properties of proteins.

Figure 4c maps the node degree distribution in the charge transfer network on to the three dimensional protein structures. The "hot" residues with significant charge transfer contributions in the protein structure are rendered in red color. Qualitatively, these "hot" residue groups with red color should be referred as somewhat "charge transfer topology associated domain (ctTAD)", which may be distinct from the structural domain in a physical model of the protein structures. Figure 4d provides the possible "hot" secondary structures in the network. As the most common building blocks for local segments of proteins, both the α-helix and β-sheet could contribute to charge transfer network. The red nodes in the loops structure and random coils are also observed in a few cases. These results illustrate non-local topology properties of the charge transfer in proteins, which may be critical to understand the underlying protein charge transfer problems.

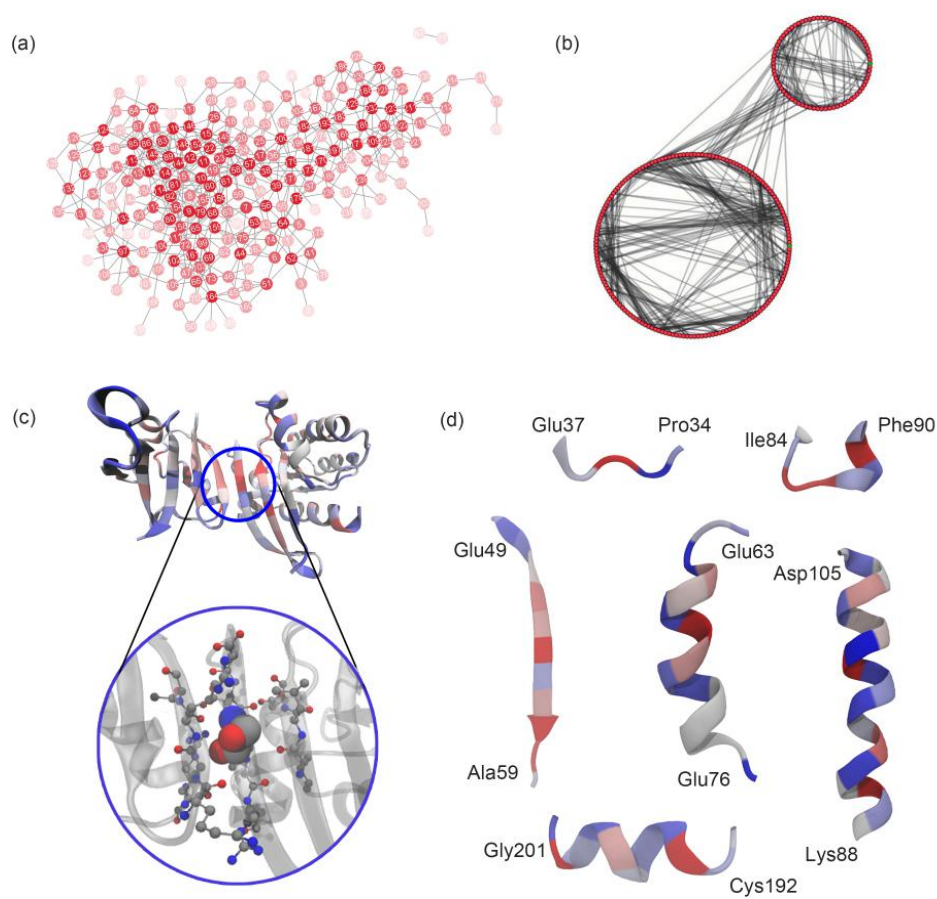

**Figure 5.** Construction of charge transfer network among residues at the resolution of 0.02 eV for a protein dimer (Ras-Raf). The graph representations of protein charge transfer network are given with CoSE (a) and circular (b) layout. (c) The node degree in the network is mapped to the three dimensional protein structures. (d) The secondary structures as building blocks with higher node degree are shown. The red color refers to the "hot" residues with large node degrees, while the blue color for "cold" residue with smaller node degree.

The D$^2$Net model is also applied to a protein dimer, which are the signal transduction cascade complex of human H-Ras protein and the Ras-binding domain of C-Raf1 (Ras-Raf). Figure 5 shows the topology properties of charge transfer network for this protein dimer, which is relatively different from charge transfer topology of the protein monomer in Figure 4. Figure 5a and 5b highlight the distinct boundaries of the charge transfer network between two sub-units in the three dimensional protein structures. The atomic details of the most critical "hot" node (residue) along with its connected nodes (residues) are also given in Figure 5c. The "hot" residues in the protein interface may be important for protein-protein interactions. Figure 5d suggests that the "hot" residues are usually spaced and insulated by "cold" residues in the same secondary structure. Therefore, these "hot" regions should prefer to undergo charge transfer among distinct secondary structures. The existence of ctTAD should also be carefully considered in the fragment based quantum chemistry calculations and future force field development, due to its strong possibility for charge flux.

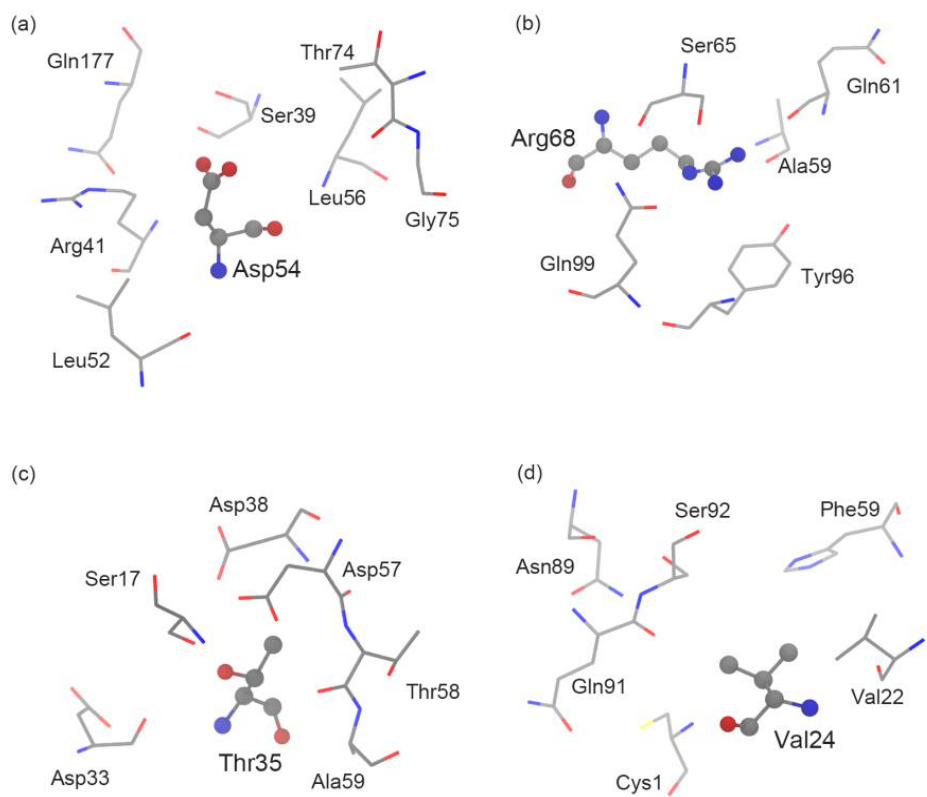

**Figure 6**. The molecular structures of several types of "hot" nodes (residues) at the resolution of 0.02 eV are extracted from the protein monomer and dimer. (a) The acidic and polar residue, (b) the basic and polar residue, (c) the neutral and polar residue, (d) the non-polar residue.

Finally, we present the atomic details of several "hot" nodes/residues along with their connected nodes in Figure 6. In the network analysis, these "hot" nodes indicate its importance role in the charge transfer reactions. Further model refinement should focus on these important motifs. As shown in Figure 6, the structural motif could also reflect the microscopic environment of the protein. The most common "hot" residues are acidic/basic and polar residues, such as aspartate and arginine in Figure 6a and 6b, which could form hydrogen bonds and even ionic bonds among amino acids. This is also consistent with our common sense in protein charge transfer reactions.[84-87] The "hot" residues may also to be neutral and polar residues, such as threonine in Figure 6c. The values of charge transfer couplings among hydrophobic or non-polar amino acids are generally low, however, the charge transfer motif with non-polar amino acids as "hot" residues is also found. Figure 6d provides the atomistic view of the valine with its surrounding amino acids. Note, the charge transfer couplings between this valine node and its connected nodes are in the range of 0.02 − 0.05 eV, while the "hot" polar residues usually exhibit larger charge transfer coupling above 0.05 eV. Thus, such "hot" non-polar residues may disappear if we adjust the resolution of the network to 0.05 eV.

**Conclusions**

The charge transfer through biological matter is one of the key steps underlying cellular energy harvesting, storage, and utilization, enabling virtually all cellular activity. As the possible structural changes will influence the electrical properties of a protein, the reasonable description of transfer couplings beyond the empirical formulas is very necessary.

In this work, we proposed the data driven network ($D^2$Net) analysis tools for understanding the electron transfer reactions in proteins. The charge transfer atlas derived from millions of amino acid conformations could be used to quickly estimate the charge transfer coupling terms, without any require of the knowledge of chemical intuition about the chemical interactions or empirical formulas. The application of $D^2$Net model revealed that the "hot" residues for charge transfer reactions can be located at the different secondary structures, i.e. α-helix, β-sheet or random coils. Therefore, the predictions of the "hot" residue or residue groups can be made from complex network analysis.

In summary, the data driven model offers us an alternative approach for efficiently identifying the critical residue or groups of residues in charge transfer reactions, which might avoid wasting computational resources. Therefore, we can understand the possible charge transfer reactions in a more explicit and more intuitionistic fashion. Future work may be possible to enumerate the most common "hot" motif that is suitable for charge transfer reactions in proteins database.

**Acknowledgements**

This work is financially supported by the National Key R&D Program of China (Grant No. 2017YFB0203405). The authors thank the support by the National Natural Science Foundation of China (Nos. 21503249, 21373124). This work is also supported by Huazhong Agricultural University Scientific & Technological Self-innovation Foundation (Program No.2015RC008) and Special Program for Applied Research on Super Computation of the NSFC-Guangdong Joint Fund (the second phase) under Grant No.U1501501.

**Supporting information**

The technical details of charge transfer integral derivation, the classification of the charge transfer atlas and the definition of molecular coordination system for amino acids are given in the supporting information.

**TOC**

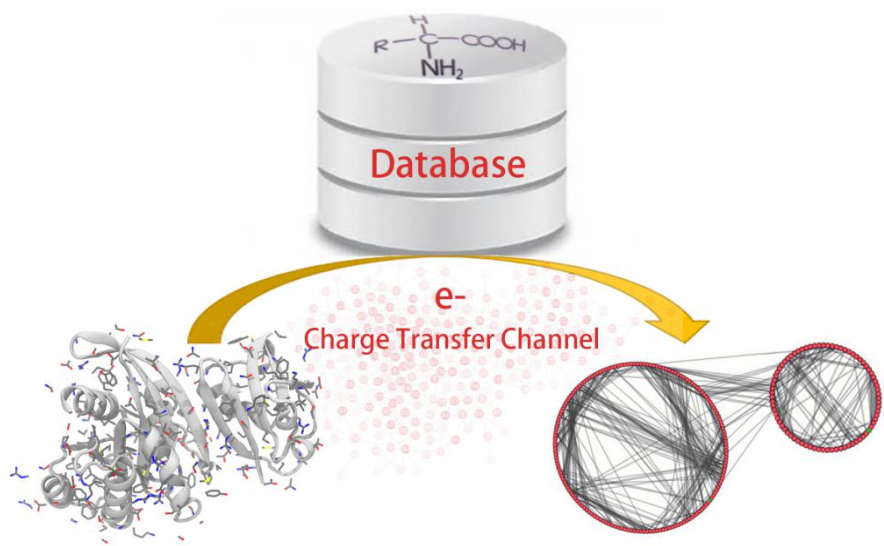

# Supporting Information

# Data Driven Charge Transfer Atlas Provides Topological View of Electronic Structure Properties for Arbitrary Proteins Complexes

Fang Liu[1], Hongwei Wang[1], Dongju Zhang[2], Likai Du[1]*

[1]Hubei Key Laboratory of Agricultural Bioinformatics, College of Informatics, Huazhong Agricultural University, Wuhan, 430070, P. R. China

[2]Institute of Theoretical Chemistry, Shandong University, Jinan, 250100, P. R. China

*To whom correspondence should be addressed.

Likai Du: dulikai@mail.hzau.edu.cn;

**Technical Details for Constructing Charge Transfer Atlas**

The procedure to extract each dimer complex has been described in the work of Singh and Thornton.[1] Two types of model subsystem were defined based on the hydrogenated amino acids: the first contained only the amino acid side chain starting from the $C_\beta$ atom ($C_\beta$ model), while the second consisted of the amino acid side chain plus the backbone $C_\alpha$ atom ($C_\alpha$ model).

Because the initial structures in the amino acid side-chain atlas contains only the coordinates of heavy atoms, the missing hydrogens were added using the *tleap* module in AmberTools package[2]. The point of cutting covalent bond was saturated with hydrogen atoms (i.e., either the $C_\alpha$ or $C_\beta$ atom). In order to eliminate any potential nonspecific interactions, the positions of hydrogen atoms were optimized for each dimer at the semiempirical PM6 method with subsequent B3LYP/6-31G* calculations as implemented in Gaussian 09 package[3]. The coordinates of the heavy atoms were kept fixed during the optimization procedure. The optimized structures are used for our subsequent construction of charge transfer atlas and network analysis.

The transfer network for the protein structure was constructed on the basis of the charge transfer atlas. First, we iteratively searched all the residues within 10 Å of each residue, which were supposed to have significant charge transfer couplings. Then, the inter-residue degrees of freedom for each residue pair were calculated. And each pairwise interaction could be represented by a vector, which was used as fingerprint to search the available charge transfer atlas with millions of side-chain pairs. We used the RMSD value as a criterion to evaluate the structural similarity between the unknown amino acid pair and the amino acid side-chain database.

If the assignment for such unknown structure is failed, an ab initio calculation of the charge transfer coupling is triggered; and this unknown structural pattern is automatically added into our charge transfer atlas for future similarity assignment. The electronic couplings for each amino acid side-chain dimer were derived from the RHF/6-31G* level according to the idea of tight-binding approximation. The quantum chemistry calculations are performed with Gaussian 09 package. After the assignment, the potential electron transfer networks for arbitrary protein are constructed.

# Derivation of Tight Binding Method for Biomolecules (bioTB)

The tight binding model is an effective approach to study complicated molecular systems with large size. The single-electron motion equation for biomolecule systems is derived according to the idea of tight-binding approximation. Here, we provide a detailed summary of the tight binding method for biomolecules (bioTB), following the previous work of Liu and co-workers[3,4]. The parameters for the on-site energy and transfer integral in the bioTB model are directly derived from ab initio calculations, which can be applied to the electronic structure calculations of biomolecules, such as DNA, proteins and *so on*.

The time independent Schrödinger equation for the many-electron problem of large biomolecules can be written as,

$$\mathbf{H}\Psi = E\Psi \tag{S1}$$

And **H** is the Hamiltonian operator for the biomolecular systems of nuclei and electrons, described by position vectors $R_a$ and $r_i$, respectively. In atomic units, the Hamiltonian for $n$ electrons and $m$ nuclei can be written as,

$$H = \sum_{i=1}^{n} -\frac{1}{2}\nabla_i^2 + \sum_{a=1}^{m} -\frac{1}{2M_a}\nabla_a^2 + \sum_{i=1}^{n}\sum_{a=1}^{m} -\frac{Z_a}{r_{ai}} + \sum_{i=1}^{n}\sum_{j<i}^{n} \frac{1}{r_{ij}} + \sum_{a=1}^{m}\sum_{b>a}^{m} \frac{Z_a Z_b}{R_{ab}} \tag{S2}$$

In Eq. S2, $M_a$ is the ratio of the mass of nucleus $a$ to the mass of an electron, and $Z_a$ is the atomic number of nucleus $a$. The Laplacian operators $\nabla_i^2$ and $\nabla_a^2$ involve differentiation with respect to the coordinates of the $i$th electron and the $a$th nucleus. The first term in the equation is the operator for the kinetic energy of the electrons; the second term is the operator for the kinetic energy of the nuclei; the third term represents the coulomb attraction between electrons and nuclei; the fourth and fifth terms represent the repulsion between electrons and between nuclei, respectively.

The kinetic energy of the nuclei can be neglected within the Born-Oppenheimer approximation. The remaining terms in Eq. S2 are called the electronic Hamiltonian or Hamiltonian describing the motion of $n$ electrons in the field of m point charges,

$$H = \sum_{i} -\frac{1}{2}\nabla_i^2 + \sum_{a,i} -\frac{Z_a}{r_{ai}} + \sum_{i<j} \frac{1}{r_{ij}} + \sum_{a<b} \frac{Z_a Z_b}{R_{ab}} \tag{S3}$$

The building block of biomolecules is simply a few repeated structure units, i.e. amino acids for proteins, nucleotides for DNA and *so on*. Thus, the many electron Hamiltonian in this physical picture can be written as,

$$H = \sum_i -\frac{1}{2}\nabla_i^2 + \sum_{i,L}\sum_{a\in L} -\frac{Z_a}{r_{ai}} + \frac{1}{2}\sum_{i,L}\sum_{j\in L, j\neq i}\frac{1}{r_{ij}} \\ + \sum_L \sum_{a,b\in L, a<b} \frac{Z_a Z_b}{R_{ab}} + \sum_{L,M}\sum_{a\in L, b\in M}\frac{Z_a Z_b}{R_{ab}} \tag{S4}$$

In the above equation, L and M refers to the repeated structure units. The last term refers to the repulsion between the nuclei, which can be considered to be constant. Any constant added to an operator has no effect on the electronic structure of the biomolecules. Therefore, we can rewritten the Eq. S4 as the sum of one-electron operator,

$$H = \sum_i \left[-\frac{1}{2}\nabla_i^2 + \sum_L V_L(i)\right] \tag{S5}$$

And

$$V_L(i) = \sum_{a\in L} -\frac{Z_a}{r_{ai}} + \frac{1}{2}\sum_{j\in L, j\neq i}\frac{1}{r_{ij}} \tag{S6}$$

Thus, the corresponding many electrons Schrödinger equation can be solved *via* the one-electron Schrödinger equation,

$$h_i\psi = \varepsilon_i\psi \tag{S7}$$

whereas, $\psi$ is the one-electron wavefunction for the biomolecules, and the one-electron Hamiltonian can be expressed as

$$h_i = -\frac{1}{2}\nabla_i^2 + \sum_L V_L(i) \tag{S8}$$

The first step to solve the the Eq. S7 is to obtain the one-electron Schrödinger equation at the non-perturbation and isolated site.

$$\left[-\frac{1}{2}\nabla_i^2 + V_L\right]\phi_l = \varepsilon_l^0 \phi_l \tag{S9}$$

whereas, $\phi_l$ is the molecular orbital of site L, and $\varepsilon_l^0$ is the corresponding orbital energy. The Eq. S9 can be solved by the Hartree-Fock or DFT method. Here, the site potential is derived from the self-consistent Hartree-Fock matrix of the amino acid side-chain dimer.

And the electronic orbital for the entire biomolecules can be expanded as the linear combination of site orbitals.

$$\psi = \sum_l c_l \phi_l \tag{S10}$$

Substituted the above equation into Eq. S8, the following equation is obtained.

$$\sum_l c_l \langle \phi_m | h | \phi_l \rangle = \varepsilon \sum_l c_l \langle \phi_m | \phi_l \rangle \tag{S11}$$

According to the tight-binding approximation, the electrons in this model are tightly bound to the site to which they belong and they should have limited interaction with states and potentials on surrounding sites. And the potential of the neighboring sites can be treated as perturbations. The tight binding Hamiltonian is usually written as,

$$H_{TB} = \sum_n \varepsilon_n c_n^+ c_n - t_{n,n+1}\left(c_{n+1}^+ c_n + c_n^+ c_{n+1}\right) \tag{S12}$$

whereas, $c$ and $c^+$ are the annihilation and creation operators for electrons or holes, $\varepsilon$ is the on-site energy and $t$ is the transfer integral between sites.

In the tight-binding approximation, the electron has limited interaction with the non-neighboring sites. Thus, we can get the following expression,

$$\langle \phi_m | h | \phi_m \rangle c_m + \langle \phi_{m-1} | h | \phi_m \rangle c_{m-1} + \langle \phi_m | h | \phi_{m+1} \rangle c_{m+1} = \varepsilon c_m \tag{S13}$$

After comparing the Eq. S12 and S13, the formulas for the on-site energy and transfer integral can be given as,

$$\varepsilon_n = \langle \phi_n | h | \phi_n \rangle = \langle \phi_n | -\frac{1}{2}\nabla^2 + \sum_L V_L | \phi_n \rangle \tag{S14}$$

$$t_{n,n+1} = -\langle \phi_n | h | \phi_{n+1} \rangle = -\langle \phi_n | -\frac{1}{2}\nabla^2 + \sum_L V_L | \phi_{n+1} \rangle \tag{S15}$$

The summation runs over all possile site L, however, only the neighboring sites are required to be considered in the tight binding approximation. Therefore, the on-site energy for site $n$ only needs the potential information of site $n$ and its neighboring site $n$-1 and $n$+1 (linear molecules), the related formula of on-site energy can be simplified as,

$$\begin{aligned}\varepsilon_n &\approx \langle \phi_n | -\frac{1}{2}\nabla^2 + V_n + V_{n-1} + V_{n+1} | \phi_n \rangle \\ &= \varepsilon_n^0 + \langle \phi_n | V_{n-1} + V_{n+1} | \phi_n \rangle\end{aligned} \tag{S16}$$

In Eq. S16, the on-site energy is not fully equal to the orbital energy of the site $n$, at least, the first neighboring sites should be considered. For non-linear molecules, such as proteins, the on-site energy for unit *n* only needs the potential information of unit *n* and its closest neighboring units **C**, and the formula of on-site energy can be simplified as,

$$\varepsilon_n \approx \langle \phi_n | -\frac{1}{2}\nabla^2 + V_n + \sum_{L \in \mathbf{C}} V_L | \phi_n \rangle$$
$$= \varepsilon_n^0 + \langle \phi_n | \sum_{L \in \mathbf{C}} V_L | \phi_n \rangle \quad (S17)$$

The use of amino acid dimers is sufficient for the construction of charge transfer coupling matrix. Thus, the transfer integral only require to the potential of site *n* and *n*+1, that is

$$t_{n,n+1} \approx -\langle \phi_n | -\frac{1}{2}\nabla^2 + V_n + V_{n+1} | \phi_{n+1} \rangle \quad (S18)$$

Because the tight binding model is corresponding to the orthogonal basis, the Löwdin method is perform to minimize the orbital overlap. And the transfer integral can be transformed to,

$$t^{eff}_{n,n+1} = \frac{t_{n,n+1} - \frac{1}{2}(\varepsilon_i + \varepsilon_j)s_{n,n+1}}{1 - s^2_{n,n+1}} \quad (S19)$$

In Eq. S19, *s* is the orbital overlap integral between sites. This transformation shows minor effects on the on-site energy and therefore ignored in our code implementation.

As two basic variables, the on-site energy and transfer integral are the diagonal and sub-diagonal elements of the tight binding Hamiltonian, which are basic physical variables in the study of DNA damage and respiration, photosynthesis and the design of molecular electronic device and charge transfer problems. The charge transfer integral describes the ability to perform the charge transfer among neighbor sites, meanwhile, the on-site energy describes the ability to move or inject an electron from a specific site.

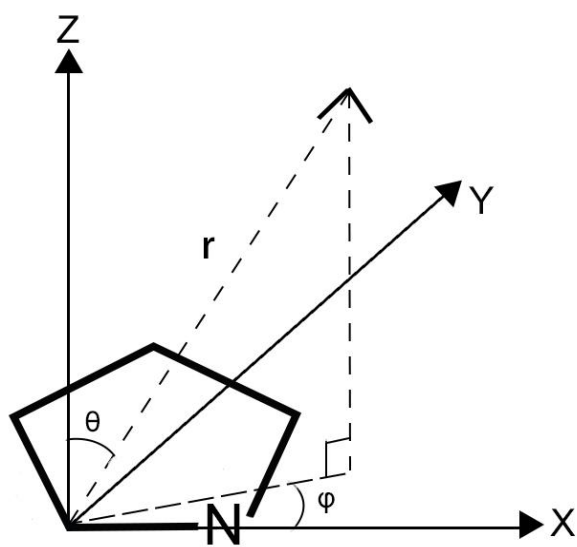

**Figure S1.** Definition of the spherical coordinate system, and the r, θ, φ is shown for the Pro side-chain as an example. Each amino acid pairs are transformed this coordinate system for subsequent clustering or RMSD calculations.

The coordinate system is defined as follows for each amino acid, which given by the Cartesian coordinates in PDB format.

**Ala**

| ATOM | 2698 | N   | ALA | F | 1 | -1.198 |  0.835 |  0.000 | 1.00 | 10.00 |
| ATOM | 2699 | CA  | ALA | F | 1 |  0.000 |  0.000 |  0.000 | 1.00 | 10.00 |
| ATOM | 2700 | CB  | ALA | F | 1 |  1.251 |  0.872 |  0.000 | 1.00 | 10.00 |

**Arg**

| ATOM | 1165 | NH1 | ARG | F | 1 | -1.149 |  0.668 |  0.000 | 1.00 | 10.00 |
| ATOM | 1166 | CZ  | ARG | F | 1 |  0.000 |  0.000 |  0.000 | 1.00 | 10.00 |
| ATOM | 1167 | NH2 | ARG | F | 1 |  1.147 |  0.667 |  0.000 | 1.00 | 10.00 |
| ATOM | 1168 | NE  | ARG | F | 1 |  0.015 | -1.329 |  0.000 | 1.00 | 10.00 |
| ATOM | 1169 | CD  | ARG | F | 1 | -1.171 | -2.175 | -0.001 | 1.00 | 10.00 |

**Asn**

| ATOM | 880 | OD1 | ASN | F | 1 | -1.082 |  0.593 |  0.000 | 1.00 | 10.00 |
| ATOM | 881 | CG  | ASN | F | 1 |  0.000 |  0.000 |  0.000 | 1.00 | 10.00 |
| ATOM | 882 | ND2 | ASN | F | 1 |  1.165 |  0.639 |  0.000 | 1.00 | 10.00 |
| ATOM | 883 | CB  | ASN | F | 1 |  0.057 | -1.514 | -0.001 | 1.00 | 10.00 |

**Asp**

| ATOM | 1324 | OD1 | ASP | F | 1 | -1.094 | 0.605 | 0.000 | 1.00 | 10.00 |
|------|------|-----|-----|---|---|--------|-------|-------|------|-------|
| ATOM | 1325 | CG | ASP | F | 1 | 0.000 | 0.000 | 0.000 | 1.00 | 10.00 |
| ATOM | 1326 | OD2 | ASP | F | 1 | 1.096 | 0.606 | 0.000 | 1.00 | 10.00 |
| ATOM | 1327 | CB | ASP | F | 1 | 0.012 | -1.515 | -0.004 | 1.00 | 10.00 |

**Cys**

| ATOM | 496 | CA | CYS | F | 1 | -1.282 | 0.835 | 0.000 | 1.00 | 10.00 |
|------|-----|----|-----|---|---|--------|-------|-------|------|-------|
| ATOM | 497 | CB | CYS | F | 1 | 0.000 | 0.000 | 0.000 | 1.00 | 10.00 |
| ATOM | 498 | SG | CYS | F | 1 | 1.515 | 0.987 | 0.000 | 1.00 | 10.00 |

**Gln**

| ATOM | 676 | OE1 | GLN | F | 1 | -1.083 | 0.593 | 0.000 | 1.00 | 10.00 |
|------|-----|-----|-----|---|---|--------|-------|-------|------|-------|
| ATOM | 677 | CD | GLN | F | 1 | 0.000 | 0.000 | 0.000 | 1.00 | 10.00 |
| ATOM | 678 | NE2 | GLN | F | 1 | 1.164 | 0.638 | 0.000 | 1.00 | 10.00 |

**Glu**

| ATOM | 892 | OE1 | GLU | F | 1 | -1.099 | 0.600 | 0.000 | 1.00 | 10.00 |
|------|-----|-----|-----|---|---|--------|-------|-------|------|-------|
| ATOM | 893 | CD | GLU | F | 1 | 0.000 | 0.000 | 0.000 | 1.00 | 10.00 |
| ATOM | 894 | OE2 | GLU | F | 1 | 1.099 | 0.600 | 0.000 | 1.00 | 10.00 |

**Gly**

| ATOM | 2644 | N | GLY | F | 1 | -1.210 | 0.806 | 0.000 | 1.00 | 10.00 |
|------|------|----|-----|---|---|--------|-------|-------|------|-------|
| ATOM | 2645 | CA | GLY | F | 1 | 0.000 | 0.000 | 0.000 | 1.00 | 10.00 |
| ATOM | 2646 | C | GLY | F | 1 | 1.262 | 0.841 | 0.000 | 1.00 | 10.00 |

**His**

| ATOM | 547 | CD2 | HIS | F | 1 | -1.085 | 0.813 | 0.000 | 1.00 | 10.00 |
|------|-----|-----|-----|---|---|--------|-------|-------|------|-------|
| ATOM | 548 | CG | HIS | F | 1 | 0.000 | 0.000 | 0.000 | 1.00 | 10.00 |
| ATOM | 549 | ND1 | HIS | F | 1 | 1.102 | 0.826 | 0.000 | 1.00 | 10.00 |
| ATOM | 550 | CB | HIS | F | 1 | 0.105 | -1.492 | -0.002 | 1.00 | 10.00 |
| ATOM | 551 | CE1 | HIS | F | 1 | 0.698 | 2.087 | -0.001 | 1.00 | 10.00 |
| ATOM | 552 | NE2 | HIS | F | 1 | -0.624 | 2.107 | -0.001 | 1.00 | 10.00 |

**Ile**

| ATOM | 1360 | CG1 | ILE | F | 1 | -1.262 | 0.870 | 0.000 | 1.00 | 10.00 |
|------|------|-----|-----|---|---|--------|-------|-------|------|-------|
| ATOM | 1361 | CB | ILE | F | 1 | 0.000 | 0.000 | 0.000 | 1.00 | 10.00 |
| ATOM | 1362 | CG2 | ILE | F | 1 | 1.259 | 0.868 | 0.000 | 1.00 | 10.00 |

**Leu**

| ATOM | 1096 | CD1 | LEU | F | 1 | -1.252 | 0.867 | 0.000 | 1.00 | 10.00 |
|------|------|-----|-----|---|---|--------|-------|-------|------|-------|
| ATOM | 1097 | CG | LEU | F | 1 | 0.000 | 0.000 | 0.000 | 1.00 | 10.00 |
| ATOM | 1098 | CD2 | LEU | F | 1 | 1.252 | 0.867 | 0.000 | 1.00 | 10.00 |

| ATOM | 1099 | CB | LEU | F | 1 | -0.003 | -0.930 | 1.213 | 1.00 | 10.00 |

**Lys**

| ATOM | 577 | CD | LYS | F | 1 | -1.259 | 0.854 | 0.000 | 1.00 | 10.00 |
| ATOM | 578 | CE | LYS | F | 1 | 0.000 | 0.000 | 0.000 | 1.00 | 10.00 |
| ATOM | 579 | NZ | LYS | F | 1 | 1.235 | 0.838 | 0.000 | 1.00 | 10.00 |
| ATOM | 580 | CB | LYS | F | 1 | -3.117 | 0.992 | 0.006 | 1.00 | 10.00 |
| ATOM | 581 | CG | LYS | F | 1 | -2.137 | 0.569 | 0.020 | 1.00 | 10.00 |

**Met**

| ATOM | 769 | CG | MET | F | 1 | -1.391 | 1.155 | 0.000 | 1.00 | 10.00 |
| ATOM | 770 | SD | MET | F | 1 | 0.000 | 0.000 | 0.000 | 1.00 | 10.00 |
| ATOM | 771 | CE | MET | F | 1 | 1.373 | 1.139 | 0.000 | 1.00 | 10.00 |

**Phe**

| ATOM | 862 | CD1 | PHE | F | 1 | -1.196 | 0.709 | 0.000 | 1.00 | 10.00 |
| ATOM | 863 | CG | PHE | F | 1 | 0.000 | 0.000 | 0.000 | 1.00 | 10.00 |
| ATOM | 864 | CD2 | PHE | F | 1 | 1.196 | 0.708 | 0.000 | 1.00 | 10.00 |
| ATOM | 865 | CE1 | PHE | F | 1 | -1.200 | 2.101 | -0.001 | 1.00 | 10.00 |
| ATOM | 866 | CE2 | PHE | F | 1 | 1.200 | 2.101 | -0.001 | 1.00 | 10.00 |
| ATOM | 867 | CB | PHE | F | 1 | 0.004 | -1.505 | -0.001 | 1.00 | 10.00 |
| ATOM | 868 | CZ | PHE | F | 1 | 0.000 | 2.796 | -0.001 | 1.00 | 10.00 |

**Pro**

| ATOM | 1525 | N | PRO | F | 1 | -1.149 | 0.910 | 0.000 | 1.00 | 10.00 |
| ATOM | 1526 | CA | PRO | F | 1 | 0.000 | 0.000 | 0.000 | 1.00 | 10.00 |
| ATOM | 1527 | CB | PRO | F | 1 | 1.200 | 0.950 | 0.000 | 1.00 | 10.00 |
| ATOM | 1528 | CD | PRO | F | 1 | -0.736 | 2.302 | 0.074 | 1.00 | 10.00 |
| ATOM | 529 | CG | PRO | F | 1 | 0.683 | 2.180 | 0.046 | 1.00 | 10.00 |

**Ser**

| ATOM | 1531 | CA | SER | F | 1 | -1.260 | 0.869 | 0.000 | 1.00 | 10.00 |
| ATOM | 1532 | CB | SER | F | 1 | 0.000 | 0.000 | 0.000 | 1.00 | 10.00 |
| ATOM | 1533 | OG | SER | F | 1 | 1.167 | 0.805 | 0.000 | 1.00 | 10.00 |

**Thr**

| ATOM | 1390 | OG1 | THR | F | 1 | -1.163 | 0.838 | 0.000 | 1.00 | 10.00 |
| ATOM | 1391 | CB | THR | F | 1 | 0.000 | 0.000 | 0.000 | 1.00 | 10.00 |
| ATOM | 1392 | CG2 | THR | F | 1 | 1.237 | 0.891 | 0.000 | 1.00 | 10.00 |

**Trp**

| ATOM | 697 | CD1 | TRP | F | 1 | -1.094 | 0.820 | 0.000 | 1.00 | 10.00 |
| ATOM | 698 | CG | TRP | F | 1 | 0.000 | 0.000 | 0.000 | 1.00 | 10.00 |
| ATOM | 699 | CD2 | TRP | F | 1 | 1.146 | 0.859 | 0.000 | 1.00 | 10.00 |
| ATOM | 700 | CE2 | TRP | F | 1 | 0.672 | 2.189 | 0.000 | 1.00 | 10.00 |

| ATOM | 701 | CE3 | TRP | F | 1 | 2.529 | 0.635 | -0.001 | 1.00 | 10.00 |
| ATOM | 702 | CH2 | TRP | F | 1 | 2.880 | 3.041 | 0.000 | 1.00 | 10.00 |
| ATOM | 703 | CZ2 | TRP | F | 1 | 1.531 | 3.291 | 0.001 | 1.00 | 10.00 |
| ATOM | 704 | CZ3 | TRP | F | 1 | 3.384 | 1.731 | 0.000 | 1.00 | 10.00 |
| ATOM | 705 | NE1 | TRP | F | 1 | -0.698 | 2.137 | 0.000 | 1.00 | 10.00 |

**Tyr**

| ATOM | 865 | CD1 | TYR | F | 1 | -1.195 | 0.716 | 0.000 | 1.00 | 10.00 |
| ATOM | 866 | CG | TYR | F | 1 | 0.000 | 0.000 | 0.000 | 1.00 | 10.00 |
| ATOM | 867 | CD2 | TYR | F | 1 | 1.194 | 0.715 | 0.000 | 1.00 | 10.00 |
| ATOM | 868 | OH | TYR | F | 1 | 0.002 | 4.170 | -0.002 | 1.00 | 10.00 |
| ATOM | 869 | CE1 | TYR | F | 1 | -1.200 | 2.107 | -0.001 | 1.00 | 10.00 |
| ATOM | 870 | CE2 | TYR | F | 1 | 1.201 | 2.106 | -0.001 | 1.00 | 10.00 |
| ATOM | 871 | CZ | TYR | F | 1 | 0.001 | 2.793 | -0.001 | 1.00 | 10.00 |

**Val**

| ATOM | 1477 | CG1 | VAL | F | 1 | -1.253 | 0.870 | 0.000 | 1.00 | 10.00 |
| ATOM | 1478 | CB | VAL | F | 1 | 0.000 | 0.000 | 0.000 | 1.00 | 10.00 |
| ATOM | 1479 | CG2 | VAL | F | 1 | 1.253 | 0.870 | 0.000 | 1.00 | 10.00 |